\documentstyle[pre,preprint,aps]{revtex}

\begin{document}


\title{Fast Domain Growth through Density-Dependent Diffusion
in a Driven Lattice Gas}

\author{Lisa K. Wickham, and James P. Sethna}
\address{Laboratory of Atomic and Solid--State Physics,\\
Cornell University,\\
Ithaca,  NY 14853--2501.}
\maketitle

\begin{abstract}
\baselineskip=15pt
We study electromigration in a
driven diffusive lattice gas (DDLG) whose continuous Monte Carlo
dynamics generate higher particle
mobility in areas with lower particle density.
At low vacancy concentrations and low temperatures,
vacancy domains tend to be faceted: the external driving force
causes large domains to move much more quickly than small
ones, producing exponential domain growth.
At higher vacancy concentrations and temperatures,
even small domains have rough boundaries:
velocity differences between domains are smaller, and
modest simulation times produce an average domain length scale
which roughly follows $L \sim t^{\zeta}$,
where $\zeta$ varies from roughly .55 at 50\% filling to roughly
.75 at 70\% filling.
This growth is faster than the $t^{1/3}$ behavior of a standard
conserved order parameter Ising model.
Some runs may be approaching a
scaling regime. A simple scaling picture which neglects velocity fluctuations,
but includes the cluster size dependence of the velocity,
 predicts growth with $L \sim t^{1/2}$.
At low fields and early times, fast growth is delayed until
the characteristic domain size reaches a crossover length which follows
$L_{cross} \propto E^{-\beta}$. Rough numerical estimates give $\beta= .37$
and simple theoretical arguments give $\beta= 1/3$.
Our conclusion that small driving forces can significantly enhance
coarsening may be relevant to the YB$_2$Cu$_3$O$_{7- \delta}$ electromigration
experiments of Moeckly {\it et al.}\cite{Moeckly}.%
\end{abstract}


\narrowtext

\section{Introduction}

The study of spinodal decomposition and coarsening in quenched Ising
models has been vigorously pursued\cite{reviews}.
Binder and Stauffer\cite{Binder}
predicted that, following a quench at $t=0$,
 the structure function of a coarsening system would grow
with a single length scale $L(t)$. Numerical studies have verified
that, when this length scale is removed from the results, the reduced
 structure factor is very nearly constant in time\cite{Lebowitz}.
 Lifshitz and Slyozov\cite{Lif-Slov} gave a further prediction:
domain size should asymptotically grow as
 $L \sim t^{1/ 3}$
for a conserved order parameter (COP) model.
Monte Carlo simulations
\cite {Barkema} have checked this result.
All of this theory describes the
equilibrium case, where nothing acts on the coarsening process
besides a thermal bath. In real systems, however, phase segregation
can be affected by several influences, including gravity, elastic stress,
 or electric fields. Such forces often push material around, instead
of preferring one phase over another. Given this wide area of potential
experimental application, it seems reasonable to ask: what happens when you
take a COP Ising model and apply a uniform force to push
particles (up spins) across the lattice?

This type of model was first introduced by Katz, Lebowitz,
and Spohn\cite{KLS}, who found that the external driving force raised $T_c$.
Subsequent research has carefully investigated the ordering phase transition
of this model.\cite{Schmittman} In addition to work which analyzed interface
roughness\cite{Leung etal} and domain shape\cite{Boal},
 one study has checked to see whether the
scaling and growth law results of the equilibrium model can carry
over to the nonequilibrium one\cite{Yeung etal}.
Such studies of the driven diffusive lattice gas (DDLG)
have almost always employed the same Monte Carlo dynamics --- those of Kawasaki
\cite{Kawasaki}. For this specific class of model, there is no barrier
to hinder particle motion along the interface between two phases, and
domains tend to elongate along the direction of the force. When considering
a directionally averaged measure of domain size, however, Yeung {\it et al.}
\cite{Yeung etal}.
found that the Kawasaki form of the DDLG showed familiar $L_{av} \sim t^{1/3}$
coarsening behavior.

The present study finds that this slow rate of
coarsening, as well as the orientation of domains, is model-dependent.
Noting that nonequilibrium problems have an inherent sensitivity to dynamics,
we have studied a DDLG with particle mobility that goes down when
the number of bonds to neighbors goes up. The resulting motion has free
diffusion of single particles across empty spaces.
(A similar bond--counting approach
was used to model electromigration of thin films on
semiconductors\cite{Ohta}, but that work did not study coarsening.)
In our model, the
external driving force bunches domains up along the field (so that they
lengthen in the transverse direction), and it can push entire domains of
vacancies across the lattice so that they sweep up other vacancies
and grow quickly. For moderate lattice fillings, the resulting domain radius
grows as $L_{av} \sim t^{\zeta}$, where $\zeta$ varies roughly from
0.4 to 0.7.
For high concentrations of particles,
the early stages of growth can be {\it exponential} in time.

When we approached the problem of coarsening in an electric field, we were
interested in fast motion of isolated particles through the middle of a
domain, rather than along an interface. Such bulk diffusion was relevant to
the electromigration studies of Moeckly, Lathrop,
and Buhrman\cite{Moeckly}. In their room temperature observations of
YB$_2$Cu$_3$O$_{7- \delta}$ thin film devices,
they found that a small electric bias ($\sim 10^3 V/$cm)
could produce macroscopic
motion of oxygen. The associated force was so tiny that it would only have
moved an oxygen atom a few lattice constants per second in a fully oxygenated
sample, where the activation energy for oxygen motion is about 1 eV
\cite{Cannelli} and
the diffusion constant is about $10^{-12} {\rm cm^2/s}$ near room temperature
\cite{Rothman}. In an oxygen depleted region, however, small forces may
have a large impact: internal friction measurements give an
activation energy of 0.1 eV for motion of a completely isolated oxygen
atom, and the chemical diffusion data of LaGraff {\it et al.}\cite {LaGraff}
suggest that the diffusion constant of YBCO can rise by more than an order
of magnitude as the oxygen in the chain plains is depleted.

To study the effects of such differences in mobility, we wanted the simplest
model that could describe a
density-dependent diffusion constant. We therefore chose a two
dimensional DDLG with
modified continuous Monte Carlo dynamics\cite{Barkema}.
Thus, we group atoms according
to their coordination $q$, increment time by an amount which increases
as the number of highly mobile atoms decreases,
and propose a move from list $q$ with probability
$$ P[q]= dt (4-q) N(q) e^{-4Jq},\eqno(1)$$
where $N[q]$ is the number of $q$
--coordinated atoms. This continuous Monte Carlo scheme
satisfies detailed balance, so the equilibrium state at $\Delta$ = 0
is that of the nearest--neighbor Ising model:
${\cal H} = -J \sum_{\langle ij\rangle} S_i S_j$. This dynamics
allows atoms with low coordination
to move quickly. It also produces a basic particle--hole
asymmetry, illustrated by the fact that isolated atoms can zip
across vacant spaces (rate 1), while isolated holes hardly move
(rate $e^{-12J}$).
We include the electric potential by accepting all proposed forward
moves, a fraction $e^{-\Delta}$ of the proposed sideways moves, and only
$e^{-2\Delta}$ of the proposed moves against the field, where $2\Delta kT$
is a local potential difference along the field\cite{timescale}.
Motivated by the YBCO experiments, we have focused much of our
attention on the limits of high particle concentration, relatively low fields,
and strong coupling to nearest neighbors (i.e. a highly
concentration--dependent mobility).

Figure \ref{snapshots} shows two of the interesting behaviors we found. In
both pictures, the black regions are vacancy clusters which move collectively
downwards as an external force pushes (white) particles up. In the
symmetry-breaking field, the vacancy blocks become short and wide
\cite{widening}. The pictures on the left (Fig. 1a and 1b) have 90\% lattice
filling. Here, isolated runaway processes dominate:
large domains move much faster than small ones and sweep up many
vacancies, thus becoming even bigger and faster. We discuss such
processes in the following section on exponential growth. The pictures
on the right have 70\% lattice filling.
Here, blocks of all length scales are moving and combining, and a mean
domain radius grows as a power in time.
Note that the late snapshot at the
bottom (Fig. 1d) resembles a scaled--up version of the snapshot at the top
(Fig. 1c).
In section III, we evaluate scaling collapses
and we construct a simple picture for domain growth in this regime.
Finally, in section IV, we will look at small driving forces.
In this limit, we find that
the early stages of growth show the $L \sim t^{1/3}$ behavior expected for
the zero field case, and then a crossover to fast growth occurs.
We interpret this crossover as the time at which the area
swept out through linear motion along the field equals the area visited
by diffusive motion, and we derive the field dependence of the domain size
at crossover.

\section{Exponential Growth}

The runaway growth of the high filling regime
is fundamentally tied to a separation of time-scales
produced by faceting.
The pictures in Figure \ref{snapshots} were generated with strong coupling
between neighbors, so atoms with two neighbors moved much more quickly
than atoms with three. In this regime, the base of each vacancy domain
tends to be
flat, with all atoms having three neighbors. After a stagnant
period, one of these strongly pinned atoms pops out of the base and leaves
behind two doubly coordinated particles. The remaining atoms then have
a lower barrier to motion. One by one, the rest of the row soon dislodges
and moves rapidly across the empty space.
Under such conditions, one would
expect the velocity of a region to be proportional to its horizontal width
(i.e. the number of ways to produce the initial break).
Figure \ref{velvs_width} shows
this behavior at low temperatures for isolated vacancy domains.
Periodically, a domain will collide with a vacancy in its path and that
will provide the initial break to move the domain through an extra row
of atoms. Again, the rate of such motion increases linearly with the width of
the domain.

Domain size is therefore a crucial factor in determining growth.
Besides moving more quickly to sweep up new vacancies, wide blocks
clear larger regions as they move. In general, we expect:
$${dn \over dt} = w\cdot \Delta v \cdot c.\eqno(2)$$
Here, $n$ is the area of the block in question, $w$ is its width,
$c$ is the concentration of
vacancies in the region ahead of it, and $\Delta v$ is the relative
velocity of the block we are describing (in comparison to that of vacancies
which it overtakes). For low temperatures and low vacancy concentrations,
small vacancy blocks will move at negligible velocities and large blocks will
move with $v \sim w$. In this regime we expect:
$${dn \over dt} \propto w\cdot w  \cdot c.\eqno(3)$$
If the width and height of a region scale similarly,
 then the above result gives:
${dn / dt} \propto n$ or $n$ growing exponentially in time.
In practice, we find that width grows more quickly than height.
This tendency should only enhance the rate of growth.

To check this prediction against our simulation, we calculated two--point
correlation functions in both horizontal and vertical directions.
For a rough measure of length-scales, we used the width at quarter max
of each of our correlation functions\cite {lengthscale}. Figure \ref{expgrowth}
 shows vertical block size, horizontal block size, and
the product of the two (a typical domain area) as a function of time.
To run the simulation efficiently enough to observe a large range of size,
we used a fast model with nearly infinite coupling (where uncoordinated
atoms always moved first, and then all singly coordinated atoms moved).
As figure \ref{90percent_3temps}
demonstrates, we found the same behavior at low temperature for
standard finite--coupling
dynamics. Note that the vertical scale on these
plots is logarithmic, so the straight line observed does indicate
exponential growth.

Notice that the runaway growth does not continue indefinitely. For very large
domains, the time required to move a full row of atoms from bottom to
top is comparable to the time between initial ``three--moves''. If the
rate of $q=2$ moves is the limiting step, then motion from each kink
in the domain can proceed independently and large blocks will approach a
terminal velocity. Crossover to this behavior will occur when the
time required to move an entire row of atoms through sequential $q=2$ moves
is approximately equal to the expected waiting time before one atom
in that row moves from a triply coordinated site. The slowdown in growth
in figure \ref{expgrowth} occurs when these two time-scales are comparable
to each other. Figure \ref{velvs_width} shows the
velocity of a vacancy domain in an empty lattice as a function of domain
width at two different temperatures. Note that the low temperature plot
is fairly linear, while the high temperature results do indeed approach a
terminal velocity. Figure \ref{90percent_3temps}
shows simulation results with a
domain roughening crossover which varies with temperature.

There is another way to produce rough domain bases and eliminate exponential
growth. In the case where vacancy clusters are constantly running into
one another,
their bases will always contain doubly coordinated atoms, and $q=2$ moves
will again be the rate limiting step. Figure \ref{near82_growth}
shows the changeover from
exponential to power-law growth as the number of vacancies increases.
Looking at configurations with $82\%$ filling, we see isolated runaway
domains whose acceleration slows as their bases become rough.
This behavior makes intuitive sense because,
in runaway growth, the size of large domains increases more quickly than
the spacing between small ones, so large domains can grow to be larger than
a typical interdomain spacing. We will find later that horizontal correlation
functions in the power-law growth regime nearly scale,
so the relationship between
domain width and horizontal domain spacing remains nearly fixed.
This is consistent with our observation that growth which starts in
the crowded, power-law regime tends to stay power-law.

\section{Power-law growth at lower fillings}

We have found that the exponential growth regime occurs for low temperatures
and low vacancy concentrations, where only a few
domains become large enough to respond strongly to the external field.
At lower particle fillings, most vacancies will join clumps soon
after coarsening begins, since most of the vacancies are connected
through atoms with single or double coordination at quench. At such
fillings, vacancy domains no longer move through a nearly stationary
sprinkling of tiny vacancy clumps. Instead, the lattice contains a
distribution of block sizes, most of which are moving steadily
in the field. Frequent collisions between domains provide sources of
fast moving atoms, so that motion is not characterized by long
waiting times with flat domain bases. Thus, we no longer expect the
velocity of a domain to be proportional to its width.

Figures \ref{collapses} and
\ref{near70growth} show results from the simulation at lower fillings.
The first, a check for dynamical scaling, gives clear evidence that
domains of all length scales are growing at similar rates.
The horizontal correlation function shows strong hints of scaling,
but the vertical correlation function has an anticorrelation dip which
grows more pronounced with time. (That is, the regions between vacancy
domains are becoming more thoroughly swept out.)
Although growth in this regime is not completely self--similar,
a scaling picture may be a useful first step towards describing
coarsening at these fillings.
Figure \ref{near70growth} shows characteristic domain area as a function
of time for 60\%, 70\%, and 80\% concentrations.
This growth is significantly faster than the
$t^{1/3}$ behavior of a zero field model. If we fit growth at each
concentration to ${\rm domain~area} \sim t^{2\zeta}$, $\zeta$ varies from
about $.65$ at 60\% filling to approximately $.75$
at 70\% and 80\% filling.

Although the behavior of our model in this moderately full regime is
complex, we have tried to piece together a simple picture which
would mimic the observations described above. We start with the question:
in a scaling regime where growth is still dominated by catch--up events,
what kind of velocity distribution would produce linear domain growth?
An elementary argument proceeds as follows: we can describe each time
in a scaling regime with characteristic horizontal and vertical length-scales
 $L_h$ and a $L_v$. In a typical
collision, the area gained by a vacancy cluster will scale with the product
of these two lengths, i.e.
$$dn \sim L_h \cdot L_v.\eqno(4)$$
A typical time between collisions will scale as the vertical length-scale
divided by the velocity difference of the two colliding domains:
$$dt \sim L_v / \Delta v.\eqno(5)$$
Together, these two results indicate that the area of a typical domain
will increase linearly in time if $\Delta v \sim 1/L_h$.

Is this scaling picture useful for understanding simulation results?
One obvious objection is that vertical correlations in our model
do not settle into a final scaling shape until late in the simulation,
and so the typical vertical spacing between domains does not scale
perfectly with the vertical height of the domains.
Also, we have neglected any enhancement in coarsening due to velocity
fluctuations, and in fact our simulation results indicate that domain areas
in this regime have somewhat faster than linear growth in time. Realizing that
our scaling picture is an approximate description, at best,
we have investigated the size dependence of domain velocities.
Recall that a large domain with several kinks in its
base should approach a terminal velocity where motion proceeds from each
kink independently. Is the dominant correction to this terminal velocity
a term of the form $v_{cor}/L_h$? Note that this the form one might
expect if the dominant correction is due to behavior at
the sides of a domain base. Figure \ref{velvs_invwidth}
shows velocity plotted against
$1/w$ for domains of various size moving through empty space at high
temperature, so that the domain bases had several kinks. We tried plotting
velocity vs. $w^x$ and find the best asymptotic linear fit for $x$= .7 to 1.2.
Figure \ref{velvs_invw_oc7} shows velocity as a function of $1/w$ in an
actual simulation run at 70\% filling. Despite poor statistics in the latter
plot, figures \ref{velvs_invwidth} and \ref{velvs_invw_oc7} together seem to
confirm that vacancy clusters in our model approach
constant velocity with $1/L_h$ corrections. Thus, our scaling picture may
provide a first step towards explaining observed growth at these fillings.

For fillings below $50\%$, we must focus on domains of particles,
instead of vacancies. These clumps of particles actually move against
the field direction while a wind of particles sweeps into them on one
side and tears particles away on the other.
Figure \ref{powers} shows preliminary simulation results
at these fillings. Domain area grows roughly as $t^{.75}$ at
20\% filling ($\zeta \approx .4$), as $t^{.95}$ at 30\% filling
($\zeta \approx .5$), and as $t^{1.1}$ at 50\% filling ($\zeta \approx .55$).
Note that growth at low fillings is dominated by the shorter time-scales
associated with motion of atoms with few neighbors. Note also that the
growth exponent $\zeta$ increases with filling. We do not at present have
an explanation for the latter effect.

\section{Low field crossover}

For high fillings and strong interparticle couplings, we have seen that large
external fields can dramatically enhance coarsening. In most experimental
applications, however, the potential difference between neighboring sites
is much less than $kT$, so it is natural to ask how weak fields affect
domain growth. In the zero field
limit, our model corresponds to standard Lifshitz-Slyozov growth with
asymptotic $L \sim t^{1/3}$ behavior. Slightly away from this limit, we
find that low fields
produce such slow coarsening
for a while, and then generate a crossover to fast growth and noticeable
anisotropy. Figure \ref{crossover} shows this behavior. First, initial
transients die
away on a time scale given by the rate of motion for $q=3$ atoms (as in
\cite{Barkema}), and
$t^{1/3}$ growth sets in. When the characteristic domain size is still much
less than $kT / 2\Delta$, growth takes off and the presence of the field
also appears in a loss of square domain symmetry. Figure \ref{cross_vsfield}
shows rough visual estimates of crossover length as a function
of field. Although this plot may include systematic error from
pinpointing a crossover in increasingly rounded curves, it strongly suggests
that the crossover length has a weak dependence on field.

To gain a physical understanding of the crossover, first note that it
represents a transition between diffusion-dominated growth, and
driven collisions produced by the external field. In this low field
regime, where the potential drop across the domain is still less than
$kT$, we can describe the driven motion with linear response theory.
We will argue that
crossover occurs when a typical block absorbs more vacancies through
concerted motion along the field than through diffusive motion.
Driven collisions should win
when the area of a circle swept out through diffusion, $\pi Dt$, is
equal to the area swept out by linear motion, i.e.
$$\pi Dt = v \cdot t \cdot w\eqno(6)$$
Note that we can replace the horizontal length scale $w$ with
a general length scale $L$, since this early growth regime is precisely
when length scales in all directions are the same.

To describe the crossover more completely, we need to know how $D$ and $v$
vary with the size of a domain in our model. For velocity, we refer back
to the section describing exponential growth, where we found $v \sim L$
whenever motion was limited by the slow rate of dislodging the first
atom from a row. To describe the variation of the diffusion constant, note
that $D \cong \omega (\Delta x)^2$, where omega is the frequency of a typical
move and $\Delta x$ is the center of mass displacement caused by such
a move. For a faceted domain, a typical move takes an atom from one rare
kink in the boundary to another. Such a move displaces the atom by a distance
of order $L$ and the center of mass by $ \Delta x \sim {1/ L}$.
Since the frequency of these moves will be proportional to $L$, we expect
that faceted domains will have $D \sim {1 / L}$.
Note that these results for $v$ and $D$ are consistent with the Einstein
relation
that should apply at such small fields:${D/{2kT}}= {v/ force}$.
(The driving force will be proportional to total charge of a block
and therefore its area.)

Plugging these results into equation $(6)$,
we find that the prediction $v \cdot L_{cross}$ = $\pi D$ becomes
$\Delta \cdot L_{cross}^2  \propto  L_{cross}^{-1}$ or
$L_{cross} \propto \Delta^{-1 / 3}$.\cite{rough}
 If this relationship correctly describes the crossover to fast
growth, then simulations with a well-developed $L \propto t^{1/3}$ growth
before crossover should follow the scaling collapse
$L(\Delta,t) = \Delta^{-1/3} {\cal L}(t\Delta)$, at least until fast growth has
taken over.
Figure \ref{crosscollapse} shows such a collapse. Considering the numerical
difficulty of achieving well-developed $t^{1/3}$ growth for a wide range
of fields, we believe an $\Delta^{-1/3}$ crossover is supported by the data; in
any case, our results strongly indicate
that length scale at crossover varies only weakly with field.

\section{Conclusions}

Thus, in our model, a reduction in external field only produces a
small increase in the minimum domain size for takeoff.
Although the departure from $t^{1/3}$ growth appears to be small at low
fillings, the field--driven takeoff at higher fillings soon leads to large
empty regions which move steadily against the
field as particles sweep quickly through their centers. For moderately
high lattice fillings, fast growth involves most of the vacancy regions.
Here, domain growth is very roughly linear in time, and horizontal correlations
come close to scaling. At very high lattice fillings and low temperatures,
fewer domains undergo significant coarsening, but those that do
have runaway, exponential growth.
In our model, explosive growth ends when domains
are large enough to have rough bases, either through thermal effects or
constant collisions. Our study has also explored the strong
impact which faceting can have on size dependence in velocities of
vacancy clusters. This is a subject with potential applications in the study of
void electromigration in small aluminum interconnects, where faceting is well
documented and voids
have been observed to move through the middle of single crystal grains
\cite{aluminum}. Most importantly, our model demonstrates that external
driving forces can be surprisingly effective in producing domain clumping
and macroscopic particle fluxes.

How such enhancement plays out in particular
experimental systems is still an open question. In the instance of Moeckly's
YBCO electromigration experiments, particle motion takes place in
the anisotropic environment of the oxygen ``chain'' planes, and our simple
model does not incorporate such inherent anisotropies. Also,
associating our model with the YBCO experiment involves abstracting
our simulated phase separation of completely filled and empty
areas to an experimental phase separation which may be less extreme.
Neutron and TEM observations of YBCO suggest that domain segregation in
these planes produces regions of more closely spaced oxygen
rows and less closely spaced rows\cite{spacing}. Still, the more open
environment of widely spaced rows does allow increased mobility\cite{LaGraff2}.
Moeckly's observations of
large oxygen-depleted regions suggest that field-induced clumping
is vital component of his experiments. Without describing the specific
characteristics of YBCO, our model provides a qualitative
check that small external driving forces can indeed facilitate the segregation
of domains with high particle mobility.

This study serves as a preliminary survey of a broad range of
interesting and potentially relevant model behaviors. An extension
to three dimensions would allow us to study the effect of a
finite-temperature roughening transition. Our study of
faceting effects should be expanded to cover other types of dynamics,
such as those which facilitate surface diffusion. Our simple picture
of coarsening in the presence of scaling clearly needs to
be modified to include departures from scaling and fluctuations
in domain velocity. And an entire regime
of low filling needs to be explored and understood.
Our work illustrates the breadth of issues involved in studying
 how a separation of time-scales due to faceting
can affect response to an external driving force.
We have demonstrated that useful approaches to this problem may be found
outside the long wavelength, late time limit.
Further work should improve our understanding of
particle motion and domain coarsening in systems which,
instead of being conveniently isolated in a thermal bath, are
knocked out of equilibrium by an external force.


We would like to thank B. Moeckly, R. Buhrman, J.~Marko, and G.~Barkema
for helpful conversations.  This work was
partly funded by the NPSC~(LKW), and the NSF under grant
DMR--91--18065~(LKW,~JPS). This research was conducted using the resources
of the Cornell Theory
Center, which receives major funding from the National Science Foundation
(NSF) and New York State.  Additional funding comes from the Advanced
Research Projects Agency (ARPA), the National Institutes of Health (NIH),
IBM Corporation, and other members of the center's Corporate Research
Institute. (Roughly 1000 IBM SP1 processor hours were used.)

%

%
\begin{figure}
\caption{
Above are snapshots from our simulation. An external field pushes white
particles up. Pictures (a) and (b) are early and late configurations
from a run with 90\% filling. Here, a few of the black vacancy regions
undergo runaway growth as they sweep down through the lattice.
Pictures (c) and (d) show early and late stages of a run with 70\%
filling. Here, blocks of all sizes are moving, and domain area growth is
close to linear in time. All pictures have $\Delta$ = 1 and $J$ = 1.5.
Animations of our simulations are available at http://www.lassp.cornell.edu/
LASSP\_Science.html (under ``Coarsening in a Driving Force'').}
\label{snapshots}
\end{figure}
\begin{figure}
\caption{
{}From bottom to top, these plots give vertical and horizontal cluster size,
and the product of the two as a function of time.
(Sizes are taken as the FWQM of two--point correlation functions,
such as those in Figure 5.)
These simulation results
correspond to 90\% filling and $\Delta = 1$. They were produced
with a fast, low temperature algorithm. Time is scaled so
that a given triply coordinated atom will wait approximately one time
unit before moving. Note that
the plot has log of cluster size vs linear time, so that a straight line
indicates exponential growth. Growth slows at late times, when
domains become so large that they have rough bases.}
\label{expgrowth}
\end{figure}
\begin{figure}
\caption{
{}From simulations with a single domain in an empty lattice:
velocity as a function of domain width for (a) $J$=2, and (b)
$J$=1. Note that the low temperature (high J) results are nearly
linear. Data are averages of 30 runs; error bars represent estimated
variations in the mean. $\Delta$ = 1.}
\label{velvs_width}
\end{figure}
\begin{figure}
\caption{
Cluster width vs time in a 90\% full lattice at three temperatures.
These plots were produced with standard
finite-temperature dynamics, and time is again scaled so all $q=3$
atoms have moved approximately once at $t=1$.
The circles have $J$ = 1.5 and show exponential growth which
crosses over to power law at late times.
The diamond plot (lowest) has $J$ =1 and
does not show well formed exponential growth. The squares have $J$= 1.25
and show an intermediate behavior. $\Delta$ =1.}
\label{90percent_3temps}
\end{figure}
\begin{figure}
\caption{
Log-linear plot of cluster area vs time. From top to
bottom the curves correspond to 80\%, 82\%, and 85\% filling.
$\Delta$ = 1.}
\label{near82_growth}
\end{figure}
\begin{figure}
\caption{
Best collapse of horizontal(a, c) and vertical(b, d)
correlation functions.
The first set of collapses is for 70\% filling
and covers a factor of 9 in time. The second set is for 60\%
and covers a factor of 7.5 in time.
$\Delta$ = 1 and $J$ = 1.5.
Note that the vertical correlation
function shows that the regions sandwiched between vacancy domains
are becoming somewhat more swept out. The later
curves are clustered together at the bottom of the collapse.}
\label{collapses}
\end{figure}
\begin{figure}
\caption{
{}From left to right, these curves correspond to 60\%, 70\%, and 80\%
filling. In these log-log plots of cluster area vs time, the asymptotic
slopes vary from about 1.3 at 60\% filling to about 1.5 at 70\% and at 80\%
filling, corresponding to $\zeta$ = .65 and $\zeta = .75$.
($J$ = 1.5 and $\Delta$ = 1.)}
\label{near70growth}
\end{figure}
\begin{figure}
\caption{
{}From simulations with a single domain in an empty lattice:
velocity plotted against inverse domain width. These results
are consistent with velocities which approach $v= v_0 - v_1/w$
at large domain size, as shown by the linear fit.
Here, $J$ = 1 and $\Delta$ = 1.}
\label{velvs_invwidth}
\end{figure}
\begin{figure}
\caption{
Velocity as a function of inverse width in an actual
simulation run with $J$ = 1.5 and $\Delta$ = 1.
Despite poor statistics, the linear fit is consistent with
a $1/w$ correction to a terminal velocity, as in Figure 8.}
\label{velvs_invw_oc7}
\end{figure}
\begin{figure}
\caption{
Preliminary results for power law in
cluster area vs time at 20\%, 30\%, 50\%, 60\%, and 70\% filling.
At each time, an approximate logarithmic derivative in cluster area vs time
is shown. (This is an effective slope over a finite time window around each
given time in a curve such as those in figure 7.)
Note that the asymptotic power increases with filling.}
\label{powers}
\end{figure}
\begin{figure}
\caption{
Crossover to fast growth for 90\% filling and $\Delta$=.1. The
circles show characteristic domain width and the squares show
domain height. The straight line shows $L \sim t^{1/3}$ growth.}
\label{crossover}
\end{figure}
\begin{figure}
\caption{
Rough visual estimates of crossover length versus external
field for several fields at 80\% filling. The straight line
shows $L \propto \Delta^{-1/3}$ for comparison.}
\label{cross_vsfield}
\end{figure}
\begin{figure}
\caption{
Collapse of cluster width vs time for several low field curves,
ranging from $\Delta= .005$ to $\Delta= .1$. The collapse follows
$L(\Delta,t) = \Delta^{-1/3} {\cal L}(t\Delta)$,
and uses runs with 80\% filling.}
\label{crosscollapse}
\end{figure}

\begin{references}
\bibitem{timescale} Note that large $\Delta$ means that more Monte
	Carlo moves are thrown away. Thus, to get a field-independent
	measure of time, we must scale the Monte Carlo time by
	$e^{-2\Delta}$.
\bibitem{widening}
Two processes tend to make domains fatter. They can
combine sideways with other domains, and the column of atoms at the side
of a domain can come loose and be swept up to the top of the domain by the
field. Widening continues until these processes are balanced by the
rate at which pile-ups of atoms at the top corners of a domain can grow
to refill a side column of particles.
\bibitem{lengthscale}
Note that this
procedure only gives a precise measure of length scale in a scaling
regime, where the basic shape of the correlation function remains
constant. In the present regime, the outer tail of the correlation curve grew
slightly more quickly than the midsection.
\bibitem{rough} Note that if domains are rough and $v$ is independent
of $L$, we expect $D \propto L^{-2}$. This still gives $L_{cross} \propto
\Delta^{-1/ 3}$.
See Gunton's article in\cite{reviews} for a discussion of various diffusion
mechanisms for rough domains.
\end{references}
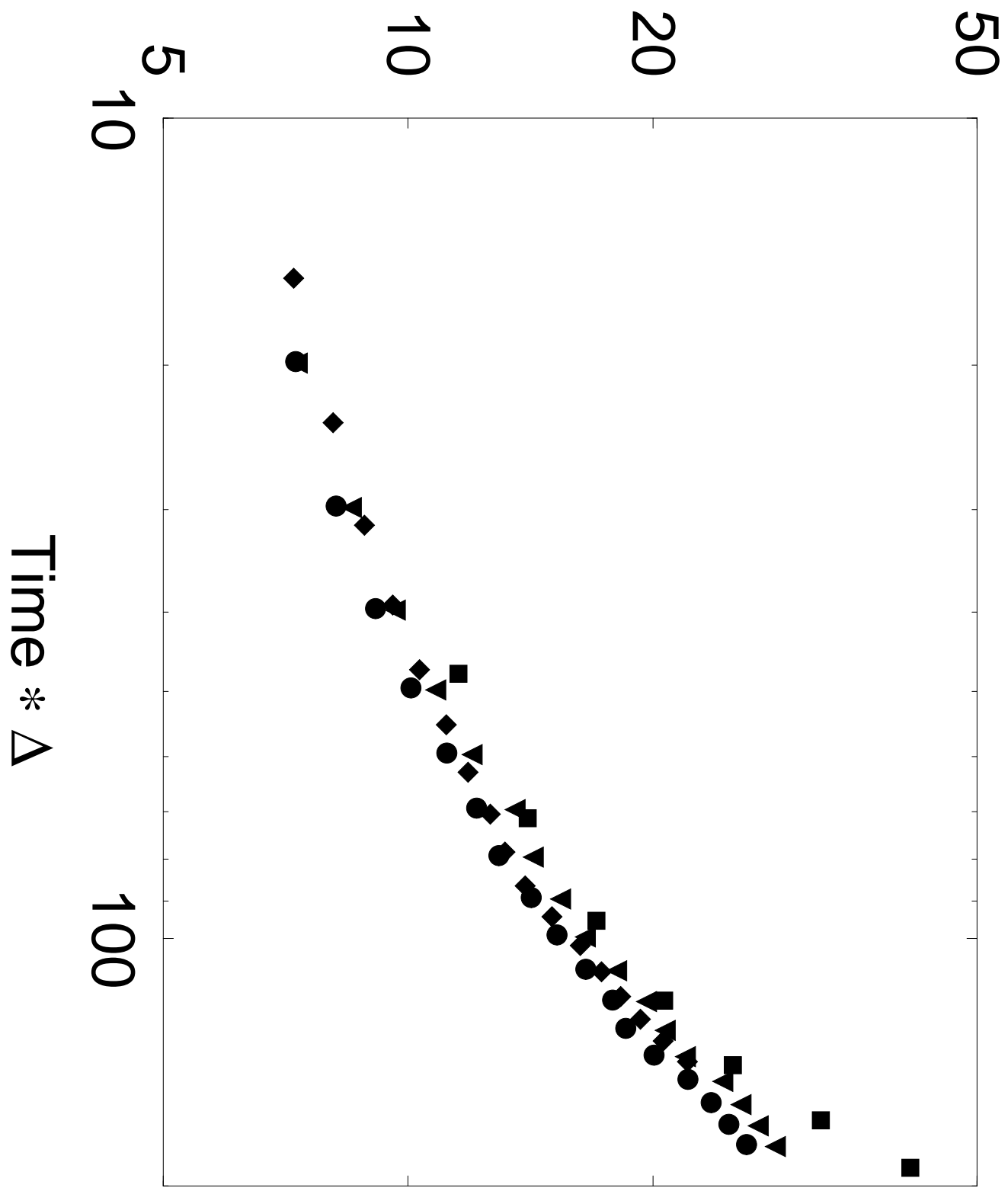
\end{document}